\documentstyle[pre,aps]{revtex}

\newcommand{\beqs}{\begin{eqnarray*}}
\newcommand{\lb}{\label}
\newcommand{\eeqs}{\end{eqnarray*}}
\newcommand{\bi}{\begin{itemize}}
\newcommand{\ei}{\end{itemize}}

\newcommand{\fr}{\frac}
\newcommand{\bt}{\beta}

\newcommand{\Ssi}{\sum_{ \{ S_i\} }}
\newcommand{\ra}{\rightarrow}
\newcommand{\siu}{ \{S_i' \} }
\newcommand{\hsu}{{\cal H}(\{S_i'\})}
\newcommand{\lan}{\langle}
\newcommand{\ran}{\rangle}
\newcommand{\hs}{{\cal H}(\{S_i\})}

\newcommand{\spv}{\sum_{<i,j>}}
\newcommand{\si}{\{S_i\}}

\newcommand{\sumCs}{\mbox{$\sum_C^*$}}
\newcommand{\Ha}{{\cal H}}
\newcommand{\desij}{\delta_{S_iS_j}}

\newcommand{\desc}{\delta_{\si,C}}

\newcommand{\jpsj}{J.Phys.Soc.Jpn.}

\newcommand{\gammaiup}{\gamma_{i\uparrow}^{\infty}}
\newcommand{\gammaidown}{\gamma_{i\downarrow}^{\infty}}
\newcommand{\gammaijpar}{\gamma_{ij}^{\parallel}}
\newcommand{\gammaijnotpar}{\gamma_{ij}^{\not\parallel}}

\begin{document}

\baselineskip=0.5cm

\title{
Percolation and Cluster Monte Carlo Dynamics for Spin Models}
\author{
V.Cataudella$^{a,b}$, G. Franzese$^a$, M. Nicodemi$^{a,b,c}$, A. Scala$^a$  
and A. Coniglio$^{a,b,c}$}
\address{$^a$Dipartimento di Scienze Fisiche, 
Universita' di Napoli I-80125 Napoli
Italy\\
$^b$ INFM - unita' di Napoli;  $^c$ INFN - sezione di Napoli}

\date{\today}

\maketitle

\begin{abstract}
A general scheme for devising efficient cluster dynamics proposed
in a previous letter [Phys.Rev.Lett. {\bf 72}, 1541 (1994)] is extensively 
discussed. In particular the strong connection among equilibrium properties of 
clusters and dynamic properties as the correlation time for magnetization
is emphasized. The general scheme is applied to a number of frustrated spin 
model and the results discussed.
\end{abstract}

\section{Introduction}

The cluster formalism introduced by Kasteleyn and Fortuin (KF)\cite{3}
and later developed by Coniglio and Klein (CK)\cite{CK} for the ferromagnetic 
Ising model has greatly enhanced the understanding 
of critical phenomena in terms of geometrical concept.
Moreover, based on such a formalism Swendsen and Wang (SW)\cite{SW87} have
introduced
a cluster dynamics which drastically reduces the critical slowing down
in Monte Carlo (MC) 
simulation of ferromagnetic spin models. 
In the SW dynamics spins belonging to the same cluster are flipped in one
step as opposed to the single spin dynamics, where spins are flipped one
at time. The efficiency of the SW dynamics stems from the fact that
in the KF-CK formalism the clusters represent correlated spins; therefore if
one spin in a cluster is flipped all the other spins in that cluster
will successively tend to flip coherently. Consequently by
flipping in one move all the spins in the same cluster results in a
much faster dynamics. 
The SW algorithm has been extended and applied efficiently to several 
unfrustrated spin models.\cite{WS90,Wolff,Niedermayer,EdwardsSokal,kawashima} 

Unfortunately, the SW dynamics based on the direct extension of the
KF-CK cluster formalism to frustrated spin systems does not show any reduction
of the relaxation times.\cite{deArc91,Cataud,leung}
The reason for that is due to the fact that now the KF-CK clusters do 
not represent any longer correlated spins.\cite{15',8b} 
Recently Kandel, Ben-Av and Domany have introduced a new type of MC cluster 
algorithm for the particular case of the fully frustrated Ising 
model\cite{BADK} which
is able to reduce the critical slowing down.
Attempts to use their algorithm to other frustrated models has been 
satisfactory in few cases \cite{wiseman,zhang} and discouraging in 
others\cite{Coddington,Kerler}.
A different approach based on the definition of quasi-frozen clusters in
spin glasses looks very promising, but the implementation of a related
cluster dynamics has not been yet explored\cite{frozen}.
More recently, based on the approach of Kandel et al., 
we have proposed a general 
criterion and a systematic procedure
to define clusters and related efficient cluster dynamics for frustrated spin 
models\cite{letter}. In particular, we have applied our general criterion
to a class of fully frustrated Ising spin models on square lattices where
the relative strength between the interactions can be varied.
For any value of the relative strength, without invoking "ad hoc" algorithms 
for each case, 
 the same general criterion
generate a Monte 
Carlo dynamics which dramatically reduce the critical slowing down.

The aim of this paper is to illustrate in more details the criterion 
proposed in Ref.\cite{letter} and apply our method to a number of frustrated 
spin models. 
In section II, III we discuss 
the extension of cluster formalism to frustrated spin models.
We stress that for the unfrustrated spin model the clusters percolate at a 
temperature $T_p$ that coincides with the critical temperature $T_c$, while
for the frustrated models $T_p$ is larger than $T_c$.
In section IV following the approach of Kandel et al.\cite{BADK}
we introduce a large variety of cluster definitions which contains as a 
particular case the KF-CK clusters. We then illustrate a procedure
to approach systematically, in successive order of approximations,
a cluster definition for which the percolation temperature $T_p$
becomes closer and closer to the critical temperature $T_c$.
In section V, VI and VII we check our procedure by comparing percolation 
quantities\cite{staufferbook,KCS}
and thermodynamic quantities on a variety of frustrated models 
using MC simulations. For a number of frustrated models without disorder
we find to second order that the clusters percolate at a temperature $T_p$
numerically indistinguishable from the critical temperature $T_c$. 
For disordered frustrated models like Spin Glass (SG) the convergence
of $T_p$ towards the SG critical temperature $T_{SG}$ is very slow.
In section VIII we implement a cluster MC dynamics based 
on the novel cluster definition.
We show that the dynamics is very 
efficient with drastic reduction of the relaxation time for those frustrated 
systems, introduced in section V-VII, for which the cluster definition
leads to $T_p\simeq T_c$.  
In the appendix we show that the cluster dynamics
fulfills detailed balance and briefly discuss the ergodicity 
problem.

\section{The cluster approach in frustrated systems.}

It is well known that he partition function of a ferromagnetic 
Ising model can be written in terms of the clusters of an equivalent
percolation model\cite{3,CK}.  A similar result can be
also obtained for Ising systems where frustration is present\cite{15'}.
The aim of this section is to recall and discuss those results which will
be useful in the following.

Let us consider the Ising Hamiltonian
\begin{equation}
\hs=-\spv J(\epsilon_{ij}S_iS_j-1)
\lb{HSG}
\end{equation}
where $\epsilon_{ij}=\pm1$ is the sign of the quenched interaction and 
$J\ge 0$ the interaction modulus. 
 The interaction 
configuration $\{\epsilon_{ij}\}$ can be a deterministic periodic structure or 
a disordered one.
Hamiltonian
(\ref{HSG}) is said to contain frustration if at least one closed path 
${\cal L}$ exists such that $\prod_{\lan i,j\ran \in {\cal L}} \epsilon_{ij} 
= -1$.

Within the CK approach\cite{CK,15',8b} we introduce "bonds" between nearest 
neighbor (nn)
spins satisfying the interaction probability $p=1-e^{-2\beta J}$.
The weight for
each given configuration of spin $\si$ and bond $C$ is given by
\begin{equation}
W_{CK}(\si,C)=p^{|C|}(1-p)^{|A|}\prod_{<i,j>\in C_F}\desij
\prod_{<i,j>\in C_A}(1-\desij)
\lb{WSG}
\end{equation}
where $C_F$ ($C_A$) is the subset of $C$ which covers 
ferromagnetic (antiferromagnetic)
bonds, i.e. bonds with $\epsilon_{ij}=1$ ($\epsilon_{ij}=-1$), and
$p=1-e^{-2\bt J}$. The product takes into account the fact that bonds can 
link only 
spins which satisfy the interaction. The clusters are defined as 
maximal sets of spins connected by bonds. 
It can be shown\cite{8b} that in this case 
the partition function becomes
\begin{equation}
Z=\sum_C \!\!\!\;^* p^{|C|}(1-p)^{|A|}2^{N(C)}
\lb{Z}
\end{equation}
where $\sum_C^*$ means that the sum is over the bond configurations 
which do not
contain frustration. Furthermore, the presence of positive and negative 
interaction implies that
the following relations can be proved
\begin{equation}
\lan S_i\ran=\lan\gammaiup\ran_{CK}-\lan\gammaidown\ran_{CK}
\lb{S_gamma_infinity_SG}
\end{equation}
and
\begin{equation}
\lan S_iS_j\ran=\lan\gammaijpar\ran_{CK}-\lan\gammaijnotpar\ran_{CK}
\lb{SS_gamma_SG}
\end{equation}
where $\gammaiup$ ($\gammaidown$) is 1 if the spin $i$ is ``up" (``down")
and belongs to an infinite cluster, otherwise it is 0, and
$\gammaijpar$ ($\gammaijnotpar$) is 1 if the spins $i$ and $j$ belong to the
same cluster and are parallel (antiparallel), otherwise it is 0, and where
$\lan...\ran$ is the usual thermodynamic
average for a fixed configuration of interaction 
$\{J\epsilon_{ij}\}$ and $\lan...\ran_{CK}$ is the average over 
spin and bond configurations weighted with (\ref{WSG}).
The percolation quantities are instead given by
\begin{eqnarray}
P_i \equiv & \lan \gamma_i^\infty\ran_{CK} & = \lan\gammaiup\ran_{CK}+
\lan\gammaidown\ran_{CK} \\
P_{ij} \equiv & \lan \gamma_{ij}\ran_{CK} & = \lan\gammaijpar\ran_{CK}+
\lan\gammaijnotpar\ran_{CK} 
\label{29}
\end{eqnarray}
where $\gamma_i^\infty=\gammaiup +\gammaidown$ and
$\gamma_{ij}=\gamma_{ij}^{\parallel} +\gamma
_{ij}^{\not\parallel}$, and $P_i$ is the probability that the spin at site $i$
belongs to the $\infty$ cluster and $P_{ij}$ is the probability that the spins
at site $i$ and $j$ are in the same cluster. 

It is clear that {\em without frustration } the relations valid in the ferromagnetic 
case are recovered with only trivial 
differences; in fact, in this case, eqs. (\ref{S_gamma_infinity_SG},
\ref{SS_gamma_SG}) become

\begin{equation}
|\lan S_i \ran|=\lan \gamma_i^{\infty}\ran_{CK}
\lb{S_gamma_infinity_a}
\end{equation}
and
\begin{equation}
|\lan S_iS_j\ran|=\lan\gamma_{ij}\ran_{CK} .
\lb{SS_gamma_a}
\end{equation}

Examples of models without frustration for which eqs. 
(\ref{S_gamma_infinity_a}) and (\ref{SS_gamma_a})
are satisfied are the antiferromagnetic Ising model
on a square lattice and an Ising model with interaction $J_{i,j}=J\sigma_i
\sigma_j$ where $\sigma_i=\pm 1$ are quenched variables.
It is easy to realize that for any fixed configuration of $\sigma_i$,
although  the interactions may be positive and negative, all the
loops are unfrustrated.

From (\ref{S_gamma_infinity_SG}) 
and (\ref{29}) it follows that unlikely the ferromagnetic case
the critical temperature does not coincide with the 
percolation temperature. In fact,
defining the critical temperature $T_c$ as the temperature at which
the Edwards-Anderson\cite{EA75} order parameter
$q_{EA}=\fr{1}{N}\sum \overline{\lan S_i\ran^2}$ vanishes and 
the percolation temperature $T_p$ as the temperature at which the
percolation probability $\rho_{\infty}=\fr{1}{N}\sum_i\overline{ P_i}$ 
vanishes, from
eq. (\ref{S_gamma_infinity_SG}) follows that $T_c<T_p$\cite{15'}.
(In the definition of $q_{EA}$ the
bar stands for the
average over the configurations of interaction $\{J\epsilon_{ij}\}$ and $N$ is
the number of spins.\cite{Notazza}) 
This result has been verified numerically for a number of
different systems as $2d$ and $3d$ spin glass\cite{deArc91,Cataud},
fully frustrated model\cite{Cataud} and frustrated $XY$ model\cite{leung}.

Similarly from eq. (\ref{SS_gamma_SG}) correlation and connectivity do not 
coincide any more. In fact two spins instead of being in the same cluster
may be parallel in one configuration and antiparallel in another (see Fig. 1).
Although these two configurations will both contribute to the connectivity 
they will interfere and strongly reduce correlations. Therefore for $T>T_c$ 
defining $g_{ij}=
\overline{\lan S_iS_j\ran}$ it follows from 
(\ref{SS_gamma_SG})
and (\ref{29}) $|g_{ij}|\leq\overline{P_{ij}}$.

\section{Generalization of the cluster approach.}

The main result of the previous section is that when frustration
is present the KF-CK clusters do not represent anymore correlated
spins. As a consequence the percolation temperature $T_p$ is higher
than the thermodynamic critical temperature $T_c$. We will generalize
now  the KF approach\cite{3} in order to define new clusters which
represent correlated spins even in the frustrated case. The new 
clusters will reduce to the usual KF-CK clusters in the unfrustrated 
case.

We will achieve this goal in two steps. First, in this section,
following the approach introduced by Ben Av at 
al.\cite{BADK,TB} (BKD), we will consider a large class of
clusters which contain as a particular case the KF-CK clusters.
Second, in the following section, we will give a criterion to choose
systematically the ``right" clusters in successive order approximations 
in such a way that pair connectness and pair correlation functions
tend to coincide. We will consider an approximation good enough when the
percolation temperature $T_p$ has approached the critical 
temperature $T_c$.

Let us consider a square lattice with ferromagnetic
and antiferromagnetic interactions\cite{nota} and let us focus our
attention on a single isolated plaquette. This can be either
unfrustrated or frustrated.\cite{-8b}
Following Ref.[\cite{BADK}], 
it is possible to generalize the KF procedure by assigning globally to each
bond 
configurations on a given plaquette or even on larger spin blocks taken as 
``elementary units" (see Fig. 2)] its own bond probability. 
Within each elementary unit or plaquette each 
bond 
configuration probability is independent on each other, and thus, 
it is not given, as in 
the CK formalism, by the product of bond probabilities of single pairs of
spins.

We consider, to begin with a specific case, a checkerboard
partition of the square lattice (see Fig.3)
and we take one of the two sets of plaquettes (the shaded or the unshaded ones)
as the set of ``elementary units" on which we make our independent choices.
We proceed further by generalizing the KF approach. Consequently
we ``dilute" the couplings on the plaquette,
replacing them with a set of new interaction configurations which contain only 
$J'\mapsto\infty$ or $J'=0$ interactions.\cite{nota_a}
We consider the generic hamiltonian (\ref{HSG}). In sec. V-VII we will
consider specific examples.
The possible interaction
configurations are shown in Fig.4.
We assign a bond weight
$w_0$ to the interaction configuration $c_0$ of the first row (no bonds), 
we assign the same weight
$w_1=w_2$...$=w_4$ to all the configurations of bonds, $c_1$...$c_4$ in the 
second row, and so on.
The symmetry of the plaquette allows us to 
choose the same weight for symmetric configurations, i.e.  
members of the same row in Fig.4.

The requirement of  the equality 
between the partition
function of the original model and that of the diluted model gives

\begin{equation}
\prod_{<i,j>\in plaq.}
e^{\bt J(\epsilon_{ij}S_iS_j-1)}
=\sum_{\alpha=0}^Mw_{\alpha}\prod_{<i,j>\in c_{\alpha}}e^{\bt 
J'(\epsilon_{ij}S_iS_j-1)}
\lb{cond_pl}
\end{equation}
where the sum is over all the possible interaction configurations on the 
plaquette ($M=15$ for the example in the Fig.4). With $<i,j>\in c_{\alpha}$ we 
mean the n.n. spins connected by $J'$ interactions in $c_\alpha$ and with 
$<i,j>\in plaq.$ the n.n. spins on the plaquette.
The previous relation can be rewritten as
\begin{equation}
\sum_{{\alpha}=0}^Mw_{\alpha}e^{-\bt\tilde{\Ha}_{plaq.}(\si,c_{\alpha})}
=e^{-\bt\Ha_{plaq.}\si}
\lb{re_cond_pl}
\end{equation}
where
\begin{equation}
\tilde{\Ha}_{plaq.}(\si,c_{\alpha})
=-\sum_{<i,j>\in c_{\alpha}}J'(\epsilon_{ij}S_iS_j-1)
\lb{Htilde_pl}
\end{equation}
and
\begin{equation}
\Ha_{plaq.}(\si)=-\sum_{<i,j>\in plaq.}J(\epsilon_{ij}S_iS_j-1)
\lb{H_pl}
\end{equation}
are the energy of the plaquette viewed as ``elementary unit" in the new 
interaction configuration $c_{\alpha}$ and in the original system.

Of course such procedure can be repeated for every Hamiltonian
that can be written as sum of elementary unit energies, i.e.
\begin{equation}
\Ha(\si)=\sum_l\Ha_l(\si) \ ,
\lb{H_sum_block}
\end{equation}
whose block $l$ we dilute and obtain the stochastic Hamiltonian
\begin{equation}
\tilde{\Ha}(\si,C)
=\sum_l\tilde{\Ha}_l( \si , c_{ \alpha (l) }  )
\lb{Htilde_sum_block}
\end{equation}
with $C=\bigcup_l c_{{\alpha} (l)}$.

The equivalence between the original model and the 
diluted one is obtained imposing (cfr. eq.(\ref{re_cond_pl}) ),
\begin{equation}
\sum_{\alpha (l)} w_{\alpha (l)}
e^{-\bt\tilde{\Ha}_l(\si,c_{\alpha (l)})}
=e^{-\bt\Ha_l(\si)}
\lb{general_cond}
\end{equation}
where the $w_{\alpha (l)}$ is 
weight with which the configuration $c_{\alpha(l)}$ on the $l$th
elementary unit occurs.

Furthermore, the partition function can be written as
\begin{equation}
Z=\Ssi \prod_le^{-\bt\Ha_l\si}=
\sum_{\si,C}\prod_l e^{-\bt\tilde{\Ha}_l(\si,c_{\alpha (l)})}
w_{\alpha(l)} \ .
\lb{Z_W}
\end{equation}

Spins that are connected by infinite strength interaction are frozen while 
the others do not interact.
Thus the dilution of the original Hamiltonian is also called 
a {\em ``freezing and deleting
operation"}. Of course we have:
\begin{equation}
e^{-\bt\tilde{\Ha}_l(\si,c_\alpha)}
=\delta_{\si,c_{\alpha (l)}}
\lb{W_delta}
\end{equation}
where $\delta_{\si,c_\alpha}$ is $1$ or $0$ depending whether or not the spin 
configurations satisfy all the $\infty$ strength interactions  in the 
interaction configuration $c_\alpha(l)$ of the $l$-th plaquette. Two spins
connected by an infinite strength interaction will be frozen in the 
configuration which satisfy the interaction.
On the entire lattice we can define:
\begin{equation}
\desc=\prod_l\delta_{\si,c_{\alpha (l)}}
=\prod_{<i,j>\in C_F}\desij\prod_{<i,j>\in C_A}(1-\desij)
\lb{gen_delta}
\end{equation}
where $C_F$ and $C_A$ are defined like in eq.(\ref{WSG}).

Let us observe that
\begin{equation}
\prod_l w_{\alpha (l)}\desc=\prod_{\alpha=0}^M 
(w_{\alpha})^{n_{\alpha}(C)}\desc
\lb{W_n}
\end{equation}
where $n_{\alpha}(C)$ is the number of elementary units 
on which we have chosen the $\alpha$th
configuration (with $\alpha=0,...M$) in the given 
interaction configuration $C$ 
(with $\sum_{\alpha} n_{\alpha}(C)
={\cal N}_u$ where ${\cal N}_u$ is the number of
elementary units).

Therefore, from eqs. (\ref{Z_W}), (\ref{W_delta}) and (\ref{W_n}), we obtain
\begin{equation}
Z=\sum_C\prod_{\alpha=0}^M
(w_{\alpha})^{n_{\alpha}(C)} \Ssi\desc
=\sum_C \!\!\!\;^* \prod_{\alpha=0}^M(w_{\alpha})^{n_{\alpha}(C)} 2^{N(C)}
\lb{Zstar}
\end{equation}
where $\sumCs$ stands for the sum over all the interaction configurations 
that are
compatible with at the least one of the possible spin configurations (i.e. 
all the unfrustrated graphs). Eq.  
(\ref{Zstar}) is the generalization of eq.(\ref{Z}) which can be recovered 
considering a pair of n.n spins as elementary unit.

Equation (\ref{Zstar}) can be also obtained following the CK approach where
the clusters are defined in the original system introducing fictitious
bonds between spins satisfying the interaction.  Given a
spin configuration $\{S_i\}$, the probability to realize a configuration of 
bonds $c_{\alpha(l)}$ on each unit $l$ is given by
\begin{equation}
P(c_{\alpha(l)}|\si)=\fr{w_{\alpha(l)}
\delta_{\si,c_{\alpha (l)}}}
{e^{-\bt\Ha_l\si}} 
\label{probability}
\lb{Pc}
\end{equation}
where $w_\alpha$ satisfy eq.(\ref{general_cond}). Due to (\ref{general_cond})
and (\ref{W_delta})these 
probabilities are normalized for any spin configuration 
$\sum_{\alpha(l)}P(c_{\alpha(l)}|\si)=1$. The Kronecker delta 
assures that the bonds are thrown only between spins satisfying the interaction.
For the entire system the weight for a given configuration of spins $\{S_i\}$
and bond configuration $C=\bigcup_l c_{{\alpha} (l)}$ is given by
\begin{equation}
W (\{S_i\},C) = \prod_l P(c_{\alpha(l)}|\si) e^{-\beta \cal H(\{S_i\})}
= \prod_l w_{\alpha(l)} \delta_{\{S_i\},c_{{\alpha} (l)}} 
\end{equation}
where (\ref{H_sum_block}) have been taken into account.
Finally from (\ref{gen_delta}) and (\ref{W_n}) we have

\begin{equation}
W (\{S_i\},C) = \prod_{\alpha =0}^M (w_\alpha)^{n_\alpha(C)} \delta_{\{S_i\},C}
.
\lb{PcW}
\end{equation}

Summing over the spin and bond configurations 
we recover eq.(\ref{Zstar}). The advantage of this approach is to make
clear that both spins and bonds can be defined in the original system where
the clusters are defined as maximal sets of spins connected by bonds.
To calculate the statistics of the CK clusters we have to generate
equilibrium spin configurations first. Then, for each equilibrium configuration
we can assign a bond configuration on each unit $l$ with probability
given by (\ref{PcW}). 

Summarizing, following the approach of BKD, 
we have defined a vast class of generalized percolation models
equivalent to our original spin model
eq.(\ref{HSG}). When one reduces the elementary 
unit to a single spin pair  one recovers KF-CK solution and that 
for larger units it is always possible to find a solution in the form
of product of KF probability $p=1-e^{-2\beta J}$, or, generally, 
solutions that are 
factorization of probabilities for a subpartition of the elementary units.

The generalization discussed does not solve automatically our problem of 
recovering the identity between cluster connectivity and spin correlation 
function, eq.(\ref{SS_gamma_a}), 
when frustration is present. However, the great freedom 
given by
eq.(\ref{general_cond}) still gives hope to find solutions
for $w_{\alpha(l)}$, in
such a way to achieve the equality (\ref{SS_gamma_a}) at least in an 
approximate way.
Then it is crucial to find a criterion to select among the many different
possibilities offered by eq.(\ref{general_cond}), those for which 
(\ref{SS_gamma_a})
holds.
This is the second step needed to achieve our aim and will be discussed in 
the next section.

\section{Conditions between correlation and connectivity.}

For each percolation model defined in the previous section
(satisfying eq.(\ref{general_cond})), it is straightforward to
generalize the relations (\ref{S_gamma_infinity_SG}) and
(\ref{SS_gamma_SG})  provided that now the average over
the spin and bond configurations has to be computed with
weights given in eq.(\ref{PcW}). In particular relation 
(\ref{SS_gamma_SG}) is also valid when we consider a 
sub-system made of a single unit $l$, namely

\begin{equation}
\lan S_iS_j\ran _l=\lan \gamma_{ij}^{\parallel}\ran _l  -
\lan \gamma_{ij}^{\not\parallel}\ran _l  ,
\lb{50a}
\end{equation}
for each $i$ and $j$ on the unit $l$. Here 

\begin{equation}
\lan ... \ran_l = \fr{\sum_{\alpha,\si}...P(c_\alpha|\si)e^{-\bt\Ha_l(\si)} }
{\sum_{\si}e^{-\bt\Ha_l(\si)}}
\lb{51}
\end{equation}
where the sum is over all  possible spin and cluster configurations
on the unit $l$ and $P(c_\alpha|\si)$ is given by (\ref{probability}).
When the quantity to average is function only of the
spin variables like $\lan s_i s_j\ran_l$, due to eq.
(\ref{probability}), eq.(\ref{51}) simplifies

\begin{equation}
\lan ... \ran_l = \fr{\sum_{\si}...e^{-\bt\Ha_l(\si)} }
{\sum_{\si}e^{-\bt\Ha_l(\si)}}
\lb{51b}
\end{equation}

Our aim is to find among the large class of solutions of 
eq.(\ref{general_cond}) a solution for $w_{\alpha(l)}$
in such a way that the equality (\ref{SS_gamma_a})
is satisfied. In this way the percolation temperature $T_p$
would coincide with $T_c$ and the clusters so identified 
will be characterized by percolation critical 
exponents equal to thermodynamical critical exponents. Since this is not a 
trivial task, we seek solutions which fulfill approximately 
eq. (\ref{SS_gamma_a}) by imposing the condition on a
sub-system made of a single unit, namely

\begin{equation}
|\lan S_iS_j\ran _l|=\lan \gamma_{ij}\ran _l .
\lb{condConiglioblockfond}
\end{equation}

To find the solution of (\ref{general_cond}) and 
(\ref{condConiglioblockfond}) may still be complicated
due to the large number of unknown ($2^n$ where $n$
is the number of edges in the unit cell). Technically the problem
can be simplified if we make use of eq. (\ref{50a}) which is
always valid if eqs (\ref{general_cond}) are satisfied.

Eq. (\ref{condConiglioblockfond}) together with
eq. (\ref{50a}) implies for each pair $ij$ in the elementary unit

\begin{equation}
\begin{array}{llcl}
\lan \gamma_{ij}^{\not\parallel}
 \ran_l = 0 & \mbox{if} & \lan S_iS_j\ran_l >0 \\
\lan \gamma_{ij}^\parallel  \ran_l = 0 & \mbox{if} & \lan S_iS_j\ran_l <0 \\
\lan \gamma_{ij}            \ran_l = 0 & \mbox{if} & \lan S_iS_j\ran_l =0 .
\end{array}
\label{55}
\end{equation}

If (\ref{55}) and (\ref{general_cond}) are satisfied, from (\ref{50a})
follows that also (\ref{condConiglioblockfond}) is satisfied. As we will
show in the next section in a specific example it is easier to impose
(\ref{general_cond}) and (\ref{55}) than to impose the equivalent
conditions (\ref{general_cond}) and (\ref{condConiglioblockfond})  .

It is clear that the larger the unit the better eq.(\ref{SS_gamma_a})
is satisfied. It also clear that there is no guarantee on how good the 
different approximations are and how fast eq.(\ref{SS_gamma_a}) is 
approached by increasing the unit size.
However,
this procedure, by increasing the size of the unit, 
allows a systematic way to improve the approximation.

Eqs.(\ref{general_cond}) and (\ref{condConiglioblockfond})
introduce a set of independent conditions whose 
number depends on the size and the symmetry of the chosen elementary unit. In 
general we are not able to know if the conditions introduced by 
(\ref{general_cond}) and (\ref{condConiglioblockfond}) have a solution and 
if it is unique. Of course only solutions such that $w_{\alpha(l)} > 0$
are acceptable and these conditions introduce further restriction.

In general, there are two possibilities. 
The first possibility is that there 
are no solutions which satisfy eqs. (\ref{condConiglioblockfond}). 
In this case we can relax eq.(\ref{condConiglioblockfond})
 by imposing
\begin{equation}
\sum_{ij} r_{ij}^k\lan \gamma_{ij}\ran_l = \sum_{ij} r_{ij}^k
|\lan S_iS_j\ran_l| 
\mbox{\hspace*{1.5cm}}k=0,1,2..k_M
\end{equation} 
where $r_{ij}$ is the distance between spin $i$ and $j$. Choosing $k_M$ in an
appropriate way it is possible to reduce the number of conditions until a 
solution is found. If $k_M \ge 2$ we believe that the solution is 
rather reasonable since the conditions of Sec. II are satisfied on the unit.

The second
possibility is that the solution is not unique. In this case we expect 
only small differences among different solutions.

To show how this scheme works, in the following sections we will analyze a
number of frustrated systems. In all cases we are always able to satisfy 
conditions (\ref{general_cond}) and (\ref{condConiglioblockfond}).

\section{A decorated Ising model.}

      In order to check our approach we have considered a decorated 
Ising model with  frustration.
Starting with an Ising model on a square lattice we introduce
between each pair of n.n. spins $S_i$, $S_j$ on the square lattice 
 two extra spins $S_k$, $S_l$  (see Fig.5) modifying the 
interaction from 
\begin{equation}
{\cal H}(S_i,S_j) = -J(S_iS_j-1)
\lb{53}
\end{equation}
 to 
\begin{equation}
{\cal H}(S_i,S_j,S_k,S_l) = -J(S_iS_j+S_iS_k+S_iS_l+S_kS_j-S_lS_j -3) \ ;
\lb{54}
\end{equation}
this generic set of four spins will be the 
elementary unit $l$. For simplicity 
in eqs. (\ref{53}) and (\ref{54}) and in the following of
this section we omit the label $l$.

     The partition function of this model is
reducible to the Ising one via a ``decimation" on spins $S_k$, $S_l$:
$\sum_{S_k,S_l} e^{-\beta {\cal H}(S_i,S_j,S_k,S_l)} =
A(J) e^{\beta \overline{J}(S_iS_j-1)}$.
The critical temperature can therefore be calculated exactly 
$T_c=2.24...$.
Monte Carlo estimation of the percolation temperature for 
the unmodified KF-CK clusters
on such a system gives $T_p-T_c\simeq 0.2$ (here and in the following 
temperature are expressed in units of $J/k_B$).
As expected the presence of frustration prevents the coincidence of 
percolation and thermodynamic properties for the unmodified 
KF-CK clusters.

We have to solve eqs. (\ref{general_cond}) and
(\ref{55}) for the unknown weights $w_\alpha$ where $\alpha$
labels the bond configurations in the elementary unit (Fig. 5).
The average in eq. (\ref{55}) is over all spin and bond configurations
with probability given by (\ref{51}) where ${\cal H}_l$
is given by (\ref{54}).

The spin correlations can be easily calculated from
(\ref{51b}) since they does not require the knowledge of the 
$w_\alpha$. We can immediately find (Fig.5)

\begin{equation}
\lan S_iS_j \ran_l > 0
\end{equation}
\begin{equation}
\lan S_iS_k \ran_l =\lan S_kS_j \ran_l =-\lan S_jS_l \ran_l =
\lan S_lS_i \ran_l >0
\end{equation}
\begin{equation}
\lan S_kS_l \ran_l = 0.
\lb{58}
\end{equation}

We can disregard bond configurations by inspection. 
For example the weight of a bond configuration which connects
$i$ and $j$ through site $l$ must be zero. In fact this bond 
configuration would correspond to $S_i$ and $S_j$ antiparallel
resulting in $\lan \gamma_{ij}^{\not\parallel}
 \ran_l > 0 $ contrary to eq. (\ref{55}). By imposing eqs. 
(\ref{55})-(\ref{58}) we
reduce the number of possible bond configurations to 12.
Furthermore 3 of them have the same connectivity properties
(i.e, they connect the same sites) as the configuration $\alpha=3$ 
(Fig. 6). Therefore they can be disregarded reducing the number
of weights  different from zero to 9.
 Eqs.(\ref{general_cond}) now read
\begin{equation}
\begin{array}{rcl}
              w_0+w_4 & = & u^4 \\
          w_0+w_1+w_4 & = & u^3 \\
         w_0+w_2+3w_4 & = & u^2 \\
 w_0+w_1+w_2+w_3+3w_4 & = & u
\end{array}
\label{dec}
\end{equation} 
where $u=e^{-2\beta J}$. The structure of such an equation system 
still allows us to chose one unknown arbitrary; in particular 
we have imposed $w_4=0$ because it provides $w_i\geq 0$ $\forall T$.

Then the solution is
\begin{equation}
\begin{array}{ll}
w_0=u^4\mbox{,                }w_1=u^3(1-u), & w_2=u^2(1-u^2) \\
w_3=u(1-u-u^2+u^3),          	      & w_4=0 \ .
\label{wdecorated}
\end{array}
\end{equation}

In order to calculate the percolation temperature of the clusters defined
by eq.(\ref{wdecorated}), $T_p$, we proceed in the following way.
Given a spin configuration $\{S_i\}$ we assign to each plaquette $l$ a bond
configuration with the probability $P(c_\alpha|\{S_i\})$ provided by
eq.(\ref{probability})
and (\ref{wdecorated}).
Then we obtain clusters defined in the entire lattice as maximal sets of spins
which are connected by bonds (CK-like cluster definition). It is, then,
possible to measure percolation quantities.

   We have estimated $T_p$ via a data
collapsing of the probability $P(T,L)$ of having a percolating cluster at
temperature $T$ in a system of size $L$ (Fig.7). 
We have simulated the model with both standard spin-flip MC dynamics and
the cluster MC dynamics which we will discuss in sec. VIII.A obtaining
indistinguishable results within our numerical precision. 
In Fig. 7 we have reported the results
obtained with the latter dynamics.
Using the scaling ``ansatz" that near
the percolation point $P(T,L) = f((T-T_p)L^{1/\nu_p})$, 
where the functional shape 
of $f$ is unknown, we have found that $T_p\simeq 2.25$
and $\nu_p\simeq 1$ consistent with the critical temperature $T_c=2.24$ and 
with the critical exponent $\nu=1$ of a ferromagnetic Ising model.

\section{The asymmetric fully frustrated model (AFF).}
Let us consider a less trivial model, the frustrated Ising model on a square 
lattice with periodic 
boundary conditions where each plaquette contains three equal ferromagnetic 
interactions $J$ and one antiferromagnetic interaction $-XJ$ ($0\le X\le 1$). 
This model interpolates between the Fully Frustrated (FF) model ($X=1$) and 
the diluted Ferromagnetic Ising model ($X=0$)\cite{note}.

If we take  $w_\alpha$ according to the definition given in Fig.8
eqs.(\ref{general_cond}) specified to this case give 
\begin{eqnarray}
w_1 + w_3 &=& u^3 \nonumber \\
w_1 + w_2 &=& u^{2+X} \nonumber \\
w_1 + 2w_2 + w_3 + w_5 + w_6 + w_7 + w_9 &=& u \label{set_1}\\
w_1 + 2w_2 + w_3 + w_4 + 2w_7 + w_9 &=& u \nonumber \\
w_1 + 3w_2 + w_4 + 2w_6 + w_8       &=& u^X \nonumber
\end{eqnarray}
where $u=e^{-2\beta J}$. 
The number of unknowns is larger than the number of 
equations.
The percolation temperatures associated to these solutions will be, in 
general, higher than the thermodynamic critical temperature. In order to have 
these two temperature as close as possible and the cluster connectivity 
representing as better as possible the spin correlation,
 as discussed in sec. IV, we can impose either
conditions (\ref{condConiglioblockfond}) or more simply eq.(\ref{55}). 
For pedagogical reasons we choose here eq. (\ref{condConiglioblockfond}). 
These are three independent equations

\begin{eqnarray}
2w_6 + 2w_7 + w_8 + 3w_9               &=& |u^X + u^3 - u - u^{2+X}|\nonumber \\
4w_2 + 2w_4 + 2w_6 + 2w_7 + w_8 + 3w_9 &=& |u^X - u^3 + u - u^{2+X}|
\label{set_2}\\
4w_3 + 2w_5 + 3w_7 + w_8 + 3w_7        &=& |u^X - u^3 -3u + 3u^{2+X}|\nonumber
\end{eqnarray}

Eqs.(\ref{set_1}) and (\ref{set_2}) have the 
solution\cite{note}
\begin{eqnarray} 
&w_1=u^3 \mbox{,\hspace{1cm}}w_2=u^2(u^X-u)&
\label{10a}\\
&w_4=u(1+u^2-2u^{X+1})& 
\label{10b}
\end{eqnarray} 
\begin{eqnarray} 
2w_5 &=& \left \{ \begin{array}{ll}    
                 u(3+u^2)-u^X(1+3u^2) & \mbox{\hspace{.2cm} } u>u^*\\
                 0                    & \mbox{\hspace{.2cm} } u<u^*
\end{array} \right \}    
\label{10d}\\ 
2w_6 &=& \left \{ \begin{array}{ll}       
                 (1-u^2)(u^X-u) & \mbox{\hspace{.2cm} } u>u^*\\
                 2w_4           & \mbox{\hspace{.2cm} } u<u^*                  
\end{array} \right \}    
\label{10c}\\ 
w_8 &=& \left \{ \begin{array}{ll}     
                0                    & \mbox{\hspace{.2cm} } u>u^*\\     
                u^X-u^3-3u+3u^{X+2}  & \mbox{\hspace{.2cm} } u<u^*
\end{array} \right \}    
\label{10e} 
\end{eqnarray} 
and $w_3=w_7=w_9=0$.
The equations (\ref{set_1}) 
and (\ref{set_2}) do not provide a unique solution. In fact a 
general solution can be found choosing $w_3$ as a free parameter. However, the
further requirement $w_\alpha \ge 0$ leads to  $w_3=0$.
The solution changes form for a
 temperature $T^*$ such that $u^* = e^{-2J/K_BT^*}$
satisfies the equation $(1+3u^2)u^X-u^3-3u=0$. This is due to the fact that at
$T=T^*$ the correlation between spins linked by the interaction $-XJ$ (let's 
call them $S_1$ and $S_2$) changes sign leading to different possible bond 
configurations. For 
example, the configuration associated to the weight $w_5$ which links 
antiferromagnetically $S_1$ and $S_2$ vanishes when $\langle S_1S_2\rangle > 
0$\cite{nota_equiv}.

It is interesting to specify the general solution for  the diluted 
ferromagnetic Ising model ($X=0$) and the FF model ($X=1$). 

In the first case ($X=0$)
it results $u < u^*$ for any temperature and we get the following non 
zero weights
\begin{equation}
w_1=u^3 \mbox{\hspace*{1cm}} w_2=u^2(1-u) \mbox{\hspace*{1cm}} 
w_4=w_6=u(1-u)^2 \mbox{\hspace*{1cm}} w_8=(1-u)^3 \mbox{\hspace*{1cm}} .
\lb{X0sol}
\end{equation}
These weights reproduce the original KF-CK solution for a plaquette; in fact 
they are 
in the form of products of $u$ and $1-u$ which are the KF-CK weights for a 
single spin pair. Since for $X=0$ the model reduces to a ferromagnetic Ising
model where some interactions have been put equal to zero, the original
KF-CK solutions reproduces the right clusters.

In the case $X=1$ it results  $u > u^*$ for any temperature and one obtains
the following non zero weights
\begin{equation}
w_1=u^3 \mbox{\hspace*{1cm}} w_4=w_5=u(1-u^2)
\lb{X1sol}
\end{equation}
which are in agreement with the cluster structure used in Ref.\cite{BADK1}.
It is worth while to note that in this limit all the bond configurations which
connect spins on opposite corners have weights equal to zero preventing the
four spins of the plaquette to belong to the same cluster even at $T=0$.

In order to check if the percolation model we have defined has the expected
properties, i.e. if $T_p = T_c$, we have studied the percolation and 
spin properties of the system with both spin-flip standard  MC dynamics
and the cluster MC dynamics which we will discuss in sec. VIII.B
obtaining indistinguishable results within our numerical precision. In 
Figs. 9 and 10 we have reported the results obtained with the latter
MC dynamics averaging at least over $6\cdot 10^4$ MC sweeps after 
discarding the first $10^4$. 

In Fig.9 we show the data collapsing for the mean cluster size 
defined as $S=\sum_s n_s s^2$ where $n_s$ is the number of clusters
of size $s$ and the sum is over finite clusters 
for three values of $X$ ($X=.5$, $X=.75$, $X=1$). For 
comparison  in Fig.10 we show, for the same $X$ values, the data collapsing
for $\chi=(\lan M^2\ran - \lan|M|\ran^2)$ where $M$ is the
magnetization. From those data we extract 
the percolation temperature, $T_p(X)$,
the critical temperature $T_c(X)$ and the critical exponents
$\gamma_p(X)$, $\nu_p(X)$, $\gamma(X)$, and $\nu(X)$
for any $X$ value. Summarizing we find
\begin{eqnarray}
T_p(0.5)\simeq 1.24 &\;\; T_p(0.75)\simeq 0.97 &\;\;  T_p(1)<0.1\\ 
\gamma_p(0.5)\simeq 1.75 &\;\; \gamma_p(0.75)\simeq 1.75 &\;\;
  \gamma_p(1)\simeq 2.0\\ 
\nu_p(0.5)\simeq 1.0 &\;\; \nu_p(0.75)\simeq 1.0 &\;\;  \nu_p(1)\simeq 1.0
\end{eqnarray}
for the percolation quantities, and
\begin{eqnarray}
T_c(0.5)\simeq 1.24 &\;\; T_c(0.75)\simeq 0.972 &\;\;  T_c(1)\simeq 0\\ 
\gamma(0.5)\simeq 1.75 &\;\; \gamma(0.75)\simeq 1.75 &\;\;  
\gamma(1)\simeq 1.51\\ 
\nu(0.5)\simeq 1.0 &\;\; \nu(0.75)\simeq 1.0 &\;\;  \nu(1)\simeq 1.0 
\end{eqnarray}
for the thermodynamic quantities. We note that
$T_p(X)\simeq T_c(X)$ within
the estimated numerical precision. Furthermore the same critical exponents
control the divergence of percolation and spin properties $\gamma_p(X)=
\gamma$(X)
and $\nu_p(X)=\nu(X)$ for all $X$ values but $X=1$. In the $X=1$ case 
we find $\nu(1)\neq\nu_p(1)=1.0$ and 
$\gamma(1)\neq\gamma_p(1)=2.0$. This result arises from the fact
that the condition $|\lan S_i S_j \ran |= \lan \gamma_{ij}\ran$
is satisfied only approximatively\cite{mistake}.
The value of $T_c(X)$, $\gamma(X)$ and $\nu(X)$ are in agreement with
the exact solution which gives  $T_c(0.5)=1.239...$, $T_c(0.75)=0.972...$, 
$T_c(1)=0$,\cite{andre}
$\nu=1$ and $\gamma=7/4$ for
$X\neq 1$, $\nu=1$ and $\gamma=3/2$ for
$X=1$.
It is worth noting that different choices of
the clusters which do not satisfy eq.(\ref{condConiglioblockfond}) as the
unmodified FK-CK clusters systematically give $T_p>T_c$ and percolation
 critical exponents
consistent with those obtained in the random 
percolation.\cite{staufferbook,KCS}

\subsection{Fractal structure.}

We have analyzed the cluster structure at $T\simeq T_p(X)\simeq T_c(X)$
\cite{letter,copanello}.
For $X\neq 1$
we have found that a typical configuration of critical clusters is a fractal 
made of a backbone and dangling bonds. The backbone is made of links 
and blobs as found in the ferromagnetic Ising model\cite{89Co} 
where the fractal dimension of the entire cluster was found to be equal
to $D=\frac{1}{2}(\frac{\gamma}{\nu}+d)$ which for dimension $d=2$
gives $D=15/8$ and the fractal dimension of the links or red bonds
was found equal to $D_R=13/24$.

For the symmetric fully frustrated model $X=1$ the structure changes
drastically. In fact all the clusters are made of self-avoiding
chains with fractal dimension given by the scaling relation
$D=\frac{1}{2}(\frac{\gamma_p}{\nu_p}+2)$.
Using the numerical result $\gamma_p\simeq 2.0$ and $\nu_p\simeq 2.0$ we
find numerically $D\simeq 2$ in agreement with the result of 
Ref.{BADK,Coddington,Kerler} that for $T=0$ predicts two percolating 
self-avoiding chains which fill up the entire system. It is interesting to note
that if the approximation could be improved such that the 
condition (\ref{SS_gamma_a}) would be exactly satisfied we would obtain
$\gamma=\gamma_p=7/4$ and $\nu=\nu_p=1$. In the plausible event that the
exact clusters are still self-avoiding chains the fractal dimension would be
$D=7/4$ identical to the fractal dimension of a self-avoiding random walk
at the $\theta$ point\cite{saw}.

\section{Ising spin glass.}
 
The most complex and interesting model in the class of spin systems 
described by the Hamiltonian (\ref{HSG})
corresponds to the case in which $\epsilon_{ij}$ is a random variable:
this introduces quenched disorder together with frustration. We
have studied the $\pm J$ Ising Spin Glass, in
the case of a square lattice with the probability
distribution
$p(\epsilon_{ij})=\kappa \delta_{\epsilon_{ij},-1} +
(1- \kappa) \delta_{\epsilon_{ij},1}$
where $\kappa$ is the concentration of antiferromagnetic interactions.
The phase diagram of such system was described by Ozeki\cite{Oze}
and exhibits at low temperature a paramagnetic-ferromagnetic transition if the
concentration of antiferromagnetic interactions $\kappa$ is enough diluted
(i.e. if $0\le \kappa \le \kappa_0$, with $\kappa_0\sim 0.1$), otherwise
there is a spin-glass transition at zero temperature
(i.e. if $\kappa_0\le \kappa \le 0.5 $).
The case $\kappa = 1/2$ correspond to the Edwards and Anderson model 
(EA).\cite{EA75}

As in the AFF case, we partition the lattice in square plaquettes of four 
spins. The system is then characterized by
two kind of plaquettes: frustrated
and unfrustrated. Frustrated plaquettes are those with one or
three antiferromagnetic interactions\cite{-8b}. 
We have analyzed such cases in
section VI where we have obtained the probabilities $P(c_\alpha|\{S_i\})$ 
for the bond configuration, $c_\alpha$, given a configuration of 
spin $\{S_i\}$ (see eq.(\ref{Pc}) and (\ref{X1sol})). 
We note that the weights given in eq.(\ref{X1sol}) for a
frustrated plaquette containing three ferromagnetic and one antiferromagnetic
interactions only depend on the interactions satisfied.
Therefore the same weights can be used for the frustrated 
plaquette containing an odd number of antiferromagnetic interactions.
We have used the unmodified KF-CK weights $p=1-e^{-2J\bt}$
for the unfrustrated ones (eq.(\ref{X0sol})). 

In order to compare the results obtained 
using the plaquette of four spins as a unit (we call this choice $2$nd 
order approximation) with those obtained
by using the single pair of spins as unit ($1$st order approximation) we
have simulated the model with both standard spin-flip MC dynamics and the 
cluster MC dynamics which we will discuss in sec. VIII.C. The percolation
quantities which we have measured, $T_p$, $\gamma_p$ and $\nu_p$, do not
depend, within our numerical accuracy, on the dynamics used.

In the case of EA model ($\kappa =0.5$), we have found a percolation 
temperature
$T_p^{(2)}\simeq 1.40 $ higher than the critical temperature $T_c=0$, but
lower than that obtained with the unmodified
KF-CK clusters where $T_p^{(1)} \simeq 1.80$.\cite{Cataud}
We have also estimated the percolation critical exponents $\nu_p$ and 
$\gamma_p$
via a data collapsing (see Fig.11), obtaining the values
$\nu_p\simeq 1.33$ and $\gamma_p \simeq 2.36$,
which are consistent with the random bond percolation values $\nu_p =4/3$ and
$\gamma_p = 43/18$.

We have also studied the region of low $\kappa$ where $T_c$ is finite and
the transition is ferromagnetic.\cite{Oze}
In Fig.12 we show $T_p^{(1)}(\kappa)$, $T_p^{(2)}(\kappa)$ and 
$T_c(\kappa)$ for $0 \le \kappa \le 0.1$ and $\kappa=0.5$. 
The values have been obtained after 
a data collapsing for the mean cluster size $S$ (eq.(19)) 
and susceptibility $\chi$,
respectively.
It's clear from Fig.12 that the percolation temperature  $T_p^{(2)}(\kappa)$ is
again lower then the one obtained for unmodified KF-CK clusters 
$T_p^{(1)}(\kappa)$, 
and higher then the 
critical temperature $T_c(\kappa)$; for values $\kappa \geq 0.1$
percolation temperatures slowly decrease reaching the $\kappa = 0.5$ value:
$T_p^{(1)}(0.5)\simeq 1.8$ and $T_p^{(2)}(0.5)\simeq 1.4$, while $T_c$ 
abruptly goes to zero.\cite{Oze}

From this analysis it comes out that neither the unmodified 
KF-CK clusters  nor our clusters are able to correctly represent 
spin correlations in Spin Glass systems. However since the percolation 
temperature $T_p^{(2)}<T_p^{(1)}$ one might expect that a systematic 
improvement can be obtained  if larger elementary units are used (Fig.2).

\section{Monte Carlo dynamics associated to percolation models}

It is possible to apply the general definition of cluster given above, 
to develop general MC cluster dynamics. 
The clusters are constructed assigning to each elementary unit one of 
the possible bond configurations according to the probability given in 
eq.(\ref{Pc}). Then the usual SW algorithm can be applied to the clusters
described above. Following Ref.\cite{BADK} it can also be 
proven that detailed balance holds (see Appendix).

\subsection{Decorated Ising model.}

We partition the original square lattice in plaquettes as described
in section V, and use the clusters defined there.
With that cluster definition we have
implemented the SW generalized cluster dynamics.

We have estimated the percolation temperature $T_p\simeq 2.25$ and
the percolation critical exponents $\gamma_p\simeq 1.77$ and $\nu_p\simeq 1$.
The values obtained are indistinguishable from
those obtained by using spin-flip MC dynamics reported in sec. V.
We have also estimated the corresponding thermodynamic quantities
$T_c\simeq 2.24$, $\gamma \simeq 1.78$ and $\nu\simeq 1.05$ which are
consistent with a ferromagnetic Ising critical point within our
numerical accuracy. 

In order to study the relaxation times we have computed the time dependent
magnetization correlations
\begin{equation}
\phi(t)=\frac{\lan |M(t)M(t')|\ran -\lan | M(t')|\ran^2}
             {\lan M(t')^2\ran -\lan |M(t')|\ran^2}
\label{phi}
\end{equation}
where $M(t)$ is the magnetization at time $t$. 
Using the new cluster dynamic we find a dramatic reduction of
the slowing down which is present for the standard SW dynamics.
The new 
dynamics has critical autocorrelation times of about $10$ MCS (Monte Carlo Step
per Spin), whose order of magnitude is comparable to those of the standard
SW dynamics for a ferromagnet of the same size at criticality.
On the contrary standard SW algorithm on Decorated Ising model shows very large 
correlation times at 
$T_c$ (see Fig.13). 
Our approach, then, reduces the critical slowing
down in this system when compared to standard SW and local spin flip dynamics.

\subsection{AFF model.}

We partition the original
lattice in elementary units as described in section VI and use bond 
configurations and the associated probabilities introduced there 
to define clusters. With such cluster definition we have then
implemented a SW generalized  cluster dynamics. 
We have estimated percolation quantities $T_p(X)$, $\gamma_p(X)$
and $\nu_p(X)$ for $X=0.5,0.75,1.0$ obtaining values indistinguishable 
from those
calculated by using spin flip MC dynamics. 
We have also computed the critical temperature $T_c(X)$ and critical
exponents $\gamma(X)$ and $\nu(X)$. 
 
In order to study the relaxation times of our SW generalized cluster
dynamics we have calculated 
the magnetization correlation function (\ref{phi}) versus Monte Carlo
sweeps at the critical temperature $T_c(X)$. It shows a dramatic reduction
for all the $X$ values which we have studied ($X=0.5,0.75,1.0$) with
respect to SW unmodified CK-KF cluster and local MC dynamics. 
For a quantitative analysis of the critical dynamic
exponent $z(X)$ at $T_c(X)$, we have calculated the integrated
autocorrelation time $\tau$ using the self-consistent procedure 
suggested in Ref.\cite{Madras} with a window equal to 6. This method
allows us to calculate $\tau$ for different system size $L$ in a
consistent way. We found a power law scaling of the form $\tau=kL^z$ 
as shown in Fig.14. The estimated values of $z(X)$ for $X=0.5,0.75$
are $z(0.5)\simeq 0.30$ and $z(0.75)\simeq 0.46$. The result
definitely shows a strong systematic reduction of the critical dynamical
exponent compared with those of standard MC dynamics. These results seem
to indicate that the criterion to let $T_p(X)$ as near as possible to
$T_c(X)$ (we have previously showed the coincidence of this two
temperature for these models), allows to individuate efficient cluster
dynamics. It is worth noting that even in the case $X=1$ when
$T_p=T_c=0$, $\nu_p=\nu=1$ and $\gamma_p>\gamma$ the cluster dynamics
exhibits a drastic reduction of the critical slowing down.
Nevertheless, our analysis suggests that, for $X=1$, it is possible improve 
further this result considering larger units as starting point of the proposed
procedure.

\subsection{$\pm$ J Ising SG model}

    The panorama is more variegated in the more complex case of Ising Spin
Glass with varying ferromagnetic interactions concentration $\kappa$. 
This model exhibits in 2d a paramagnetic-ferromagnetic 
transition for $0 \leq \kappa \leq 0.1$ 
and a 
spin-glass transition for $0.1 \leq \kappa \leq 0.5$.
Analogously to the other presented
cases the cluster dynamic is realized by using the clusters defined in 
sec.VII.
For frustrated plaquettes we have used the probabilities calculated
in sec.VI for $X=1$ (eqs. (\ref{X1sol}) and (\ref{Pc})), while for 
unfrustrated plaquettes we have used unmodified KF-CK clusters
(eqs. (\ref{X0sol}) and (\ref{Pc})).

To check our simulation, we measured thermodynamic functions as energy $E$
and specific heat $C_v$.\cite{RMC} They reproduce
known data in literature up to a temperature, $T_f$, under which our 
MC cluster dynamics freezes. We have also estimated $T_p$, $\gamma_p$
and $\nu_p$ obtaining good agreement with the values obtained 
by using spin flip MC dynamics.

We have already noted that in Ising Spin Glasses the percolation temperature 
of unmodified KF-CK clusters, $T_p^{(1)}(\kappa)$, is higher than the 
percolation temperature, $T_p^{(2)}(\kappa)$, of the clusters defined
in section VII: we have
\begin{equation}
T_p^{(1)}(\kappa) \ge T_p^{(2)}(\kappa) \ge T_c(\kappa)
\end{equation}
where $T_c$ is the thermodynamical critical temperature.
The equality holds only for $\kappa =0$ and 1 (ferromagnetic and 
antiferromagnetic case respectively).

It is possible to summarize the
results in this way: as $\kappa$ departs from the ferromagnetic Ising model 
$\kappa=0$, 
relaxation times for temperature close to the critical temperature 
$T_c(\kappa)$ get longer,
and in the region where ferromagnetic phase disappears
($0.1\le \kappa \le 0.5$), they become extremely long, even if always 
shorter than those of both standard SW cluster dynamics 
(unmodified KF-CK clusters) and local spin flip dynamics.

Along the paramagnetic-ferromagnetic transition line
(i.e. at $T_c(\kappa)$ with $0 \le \kappa \le 0.1 $) we have estimated the
critical autocorrelation time $\tau(\kappa)$, defined as the time to reduce
 square magnetization correlation to 1/10 of its value at $t=0$.
These results are shown in Fig.15 for a square lattice of size $L=32$.

In the region where the ferromagnetic 
phase disappears ($0.1\le \kappa \le 0.5$)
and the SG transition takes over at $T_{SG}=0$, 
our simulations get worse. We have studied for the case $\kappa=0.5$
the following relaxation function 

\begin{equation}
q(t)=\fr{1}{N}\sum_i\lan S_i(t_0)S_i(t+t_0)\ran \ .
\label{qt}
\end{equation}
 as a function of time (Monte Carlo Step) for systems whose size is $L=
80,90,100$. Due to  very long autocorrelation times  we were able to
 perform simulation
up to $T_f\simeq 0.8$. Averages in eq.(\ref{qt}) were taken over $1 \div 4 
\cdot 10^4$ MCS
discarding  the first $5 \cdot 10^3$ MCS.
We observed that relaxation time of our cluster dynamics above $T_f$ 
is at least one order of magnitude lower than that of a standard spin flip 
dynamics (see Fig.16 and, for comparison, Ref.\cite{McM}).

In conclusion in the case of spin glass we see that to a lowering of the 
difference $|T_c - T_p|$ corresponds a reduction of the relaxation times.
However, there are indications that such a reduction exists only for $T>T_f$.
Our results suggest that 
taking a partition of the lattice made by larger ``elementary"
 units the procedure
we have discussed define clusters whose percolation temperature is closer to
the critical temperature of the original spin model. 
The associated cluster dynamics is expected to be  characterized by shorter 
autocorrelation times. Work is in progress in this direction.

\section{Summary.}

In this paper we have discussed a general scheme for devising efficient MC 
cluster dynamics for spin models. The scheme is based on three main steps. The 
first one consists in choosing a partition of the lattice into "elementary 
units". Then, using a method first introduced by  Kandel, Ben-Av and 
Domany\cite{BADK} which is based on independent choices on each elementary 
unit, it is possible to define a vast class of cluster models whose Free Energy
is identical to the original spin model. 
Finally, among the many cluster models it is 
possible to choose the one which satisfies at the best the equality between
cluster connectivity and spin correlation (cfr. eqs.(\ref{SS_gamma_a}) and 
(\ref{condConiglioblockfond})). This procedure defines clusters which can be 
used to implement a MC cluster dynamics. We have applied this method to a number
of $2d$ frustrated spin models taking as elementary unit a single plaquette.
We show that every time $T_c\simeq T_p$
(i.e. the thermodynamic critical temperature of the spin model $T_c$ is equal 
to the percolation temperature for the equivalent percolation model) the 
associated MC cluster dynamics is characterized by very small autocorrelation 
times and a critical dynamic exponent $z$ much smaller than the one obtained 
in local (Metropolis) MC dynamics. 
We have also shown that in the more complex case of a spin glass 
where disorder is added to frustration the percolation temperature $T_p$ 
results larger than spin glass temperature $T_{SG}$ and the percolation 
critical exponent are consistent with random bond percolation exponents. 
However we 
see that in this case the percolation temperature can be decreased up to 
$T_p\simeq 1.4$ using a lattice partition based on a single plaquette. This 
result suggests that taking larger elementary units, like the ones in 
Fig.2d, $T_p$ can be further reduced. Then, we still find a 
lowering of the autocorrelation time for $T>T_f\simeq 0.8$ but, this time, 
there are no indication of a lower $z$ compared to standard Metropolis 
dynamics.

In conclusion our procedure allows for a systematic decrease of the 
autocorrelation times and, therefore, may serve as a general framework for 
the development of efficient MC dynamics in frustrated spin models. 

\section*{Appendix}

The aim of this appendix is to show 
that the MC dynamics defined in sec. VIII satisfies detailed balance. 
Following Ref.\cite{BADK} we show that
provided the mapping from $\Ha$ to $\tilde{\Ha}$ 
(see eq.(\ref{general_cond})), a MC dynamic 
which verifies {\em detailed balance principle} for $\tilde{\Ha}$ verifies 
it also for $\Ha$. In particular after executing a {\em freezing and deleting} 
operation on the original spin system, 
we can implement with the built clusters a cluster 
dynamic based for example on random flipping of independent clusters as in 
SW procedure. This is possible because such a dynamics certainly 
satisfies the {\em detailed balance principle} and is generally 
ergodic at finite temperature. 

To prove that detailed balance is respected, let us make the following
preliminary considerations. 
We can rewrite the relation (\ref{general_cond}) as
\begin{equation}
\sum_{\alpha (l)}\fr{w_{\alpha (l)}
e^{-\bt\tilde{\Ha}_l(\si,c_{\alpha (l)})}}{e^{-\bt\Ha_l\si}}=1 \ .
\lb{norm}
\end{equation}
This is the normalization condition for the conditioned probability to have 
the bond configuration $c_{\alpha (l)}$ on the 
$l$-th elementary block, given the spin configuration $\si$ on the
system, i.e. the (\ref{norm}) expresses the normalization condition for the
\begin{equation}
P(c_{\alpha(l)}|\si)=\fr{w_{\alpha (l)}
e^{-\bt\tilde{\Ha}_l(\si,c_{\alpha (l)})}}{e^{-\bt\Ha_l\si}} \ .
\lb{P}
\end{equation}
Since the choices on the elementary block are independent, the
probability of the bond configuration $C$ on the whole system, given the
spin configuration $\si$, is the product 
$P(C|\si)=\prod_lP(c_{\alpha (l)}|\si)$.

To obtain detailed balance principle, we must impose the following 
condition:
\begin{equation}
e^{-\bt\hs}T(\si\ra\siu)=e^{-\bt\hsu}T(\siu\ra\si)
\lb{det_bal}
\end{equation}
where $T(\si\ra\siu)$ is the transition probability from state $\si$ to 
state $\siu$.

By definition $T(\siu\ra\si)$ may then be written as
\begin{equation}
T(\siu\ra\si)=\sum_CP(C|\siu)\tilde{T}_C(\siu\ra\si)
\lb{T}
\end{equation}
where $\tilde{T}_C(\siu\ra\si)$ is the transition probability associated to
the dynamic
that we use on the dilute system with the Hamiltonian $\tilde{\Ha}$ (for 
example this dynamic may be the simple random flipping of independent clusters).
Generally let's suppose that the $\tilde{T}_C(\siu\ra\si)$ respects the
detailed balance principle, i.e.
\begin{equation}
e^{-\bt\tilde{\Ha}(\siu,C)}\tilde{T}_C(\siu\ra\si)=
e^{-\bt\tilde{\Ha}(\si,C)}\tilde{T}_C(\si\ra\siu) \ .
\lb{Ttilde_det_bal}
\end{equation}

Therefore, from the (\ref{T}) and the definition of $P(C|\si)$, we have
\begin{equation}
T(\siu\ra\si)=
\sum_C\left(
\prod_l\fr{w_{\alpha (l)}e^{-\bt\tilde{\Ha}_l(\siu,c_{\alpha (l)})}}
{e^{-\bt\Ha_l\siu}}
\right)\tilde{T}_C(\siu\ra\si) \ ,
\lb{1}
\end{equation}
that, using the (\ref{Ttilde_det_bal}), becomes

\begin{equation}
T(\siu\ra\si)=
\sum_C\left(
\prod_l\fr{w_{\alpha (l)}e^{-\bt\tilde{\Ha}_l(\si,c_{\alpha (l)})}}
{e^{-\bt\Ha_l\siu}}
\right)\tilde{T}_C(\si\ra\siu) \ .
\lb{2}
\end{equation}
Now, multiplying and dividing for $e^{-\bt\hs}$, we obtain
\begin{equation}
T(\siu\ra\si)=
\fr{e^{\bt\hs}}{e^{-\bt\hsu}}\sum_C\left(
\prod_l\fr{w_{\alpha (l)}^le^{-\bt\tilde{\Ha}_l(\si,c_{\alpha (l)})}}
{e^{-\bt\Ha_l\si}}
\right)\tilde{T}_C(\si\ra\siu) \ ,
\lb{3}
\end{equation}
and by (\ref{T}) and the definition of $P(C|\si)$, we get  
\begin{equation}
T(\siu\ra\si)=\fr{e^{\bt\hs}}{e^{-\bt\hsu}}T(\si\ra\siu) \ .
\lb{4}
\end{equation}
This expression is the (\ref{det_bal}) and therefore the validity of the 
principle is demonstrated\cite{BADK}. 

Summarizing, the main assumptions underling this proof consist in 
supposing that 
we can write the Hamiltonian as sum on elementary blocks $\Ha=\sum_l\Ha_l$, 
the choices on the elementary blocks are independent from each other and 
we are using a dynamics for the dilute system $\tilde{\Ha}$ that respects the
detailed balance principle. 

In particular a generalized cluster MC dynamics may be implemented with the 
following steps: 
individuate the clusters with ``freezing and deleting" (i.e. to map 
$\Ha$ into $\tilde{\Ha}$); random flip of such clusters 
(this move certainly verifies detailed balance because clusters are 
not interacting in $\tilde{\Ha}$), and iterate. 

About the ergodicity we can say that the cluster dynamics here described is
certainly ergodic for every finite temperature, because the
probability to go from a given spin configuration to any other is always
different from zero for every non zero temperature. 
In general ergodicity 
at zero temperature is difficult to prove: it must be checked specifically 
in each particular case. 
Nevertheless it is possible to guarantee ergodicity also in such extreme 
conditions, alternating cluster moves with a dynamics that certainly
is ergodic at this temperature, without changing the qualitative features 
of the cluster dynamics\cite{BADK}.

In conclusion we have proven that adopting the proposed mapping of 
Hamiltonians, and in particular for our general definition of 
clusters, it is possible to develop MC dynamics which verify detailed 
balance principle.

\section*{Figure Captions}

\begin{enumerate}
\item[Fig.1] A schematic example of two spin and bond
configuration.  The spin at site $1$ belongs to the infinite cluster
in both configurations (a,b) with different orientations. While
both configurations give positive contributions to the percolation 
probability $P_1$ they give opposite configurations to the magnetization
$\lan S_1\ran$.
Similarly the spins at sites $1$ and $4$ are connected in both configurations 
($\gamma_{14}=1$). However, they are parallel ($\gamma_{14}^{\parallel}=1$)
in a) and antiparallel ($\gamma_{14}^{\not\parallel}=1$) in b). Therefore
both configurations give a positive contribution to the pair connectness
function $p_{14}$ but opposite contribution to the pair correlation function
$\lan S_1S_4 \ran$. 

\item[Fig.2] Examples of possible elementary units partitioning a 
square lattice. The unmodified KF-CK clusters (see text) are constructed
 starting from elementary unit b). The clusters discussed in sec. V, VI
and VII make use of elementary unit c).

\item[Fig.3] A checkerboard partition of a square lattice. To cover all 
the lattice, we can choose the
shaded plaquettes, or the unshaded ones, as set of elementary units.

\item[Fig.4] The possible bond configurations on the square plaquette: full
lines are infinite interactions or present bonds, while zero interactions are 
not marked. The label $\alpha$ is the index 
of configuration $c_{\alpha}$ and of the corresponding statistical weight 
$w_{\alpha}$. As an example the configuration $\alpha=0$ has no bond present.

\item[Fig.5] The Decorated Ising model described in the text: each pair of 
interacting spin $S_i$, $S_j$ in a square lattice  is 
decorated by spins $S_k$, $S_l$. Full lines are ferromagnetic 
interactions, dashed lines are antiferromagnetic interactions.

\item[Fig.6] The bond configurations of the decorated Ising model (sec. V)
whose weights are different from zero (eq.57). We assign the same weight
to all the elements belonging to the same group. In the figure each group
is identified by a curl bracket. The conventional representation of 
interactions is the same as in the Fig.5.

\item[Fig.7] Decorated Ising model in two dimension (see sec. V): 
data collapsing for the 
probability $P(T,L)$ of having a percolating 
cluster at temperature $T$ in a systems of size $L=10,20,30$ (for each size the 
number of system spins is $5\times L^2$). The data have been obtained by
 using the cluster dynamics discussed in sec. VIII.A.

\item[Fig.8] Bond configurations for the AFF model (sec. VI). The 15
 possible bond configurations are grouped, by symmetry, in 9 groups.
To each element of a group is assigned the same weight $w_\alpha$ (cfr. eqs.
(59) and (60)). In the figure we show only one element for each group; the other
elements can be obtained trivially conserving the number of ferromagnetic
and antiferromagnetic bonds. In the brackets it is shown the number of
bond configurations belonging to the specific group. Full (dashed) lines
are bonds between ferromagnetically (antiferromagnetically) interacting
spins.

\item[Fig.9] The data collapsing for mean cluster size $S$ of AFF models 
for systems with $X=0.5,0.75,1.0$ and number of sizes $L$. The data have 
been obtained by using the cluster dynamics discussed in sec. VIII.B.

\item[Fig.10] The data collapsing for susceptibility for the AFF model,
with the same values of $X$ and $L$ as in  Fig.9.

\item[Fig.11] The data collapsing for
mean cluster size $S$ for EA Spin Glass model for systems with size 
$L=48,64,80,100$. Critical exponents and temperature are also reported.

\item[Fig.12] The critical temperature $T_c(\kappa)$ , the percolation 
temperature $T_p^{(1)}(\kappa)$, for unmodified FK-CK clusters, and the 
percolation temperature $T_p^{(2)}(\kappa)$, for the clusters discussed in 
sec. VII, versus the 
antiferromagnetic interactions concentration $\kappa$ for an Ising Model with
variable antiferromagnetic interaction concentration. The arrows show the 
values of $k=0.5$.
The data reported have been extracted by data collapsing of mean cluster size,
$S$, and susceptibility $\chi$ (see text). The cluster dynamics used is 
described in sections VIII.C.

\item[Fig.13] Correlation function, $\phi(t)$, as function of time
(MC steps per spin) for the decorated Ising model introduced in sec. V
at $T=1.6$ and for a system size $L=32$.
Two MC dynamics are compared: unmodified FK-CK cluster dynamics
(dashed line) and the cluster dynamics introduced in sections VIII.A 
(full line).

\item[Fig.14] The relaxation time $\tau$ versus system size $L$ for the 
AFF model (sec. VI) with $X=0.5,0.75$. Assuming a power law scaling 
$\tau=kL^z$ 
the estimated values of $z(x)$ are $z(0.5)=0.30$ and $z(0.75)=0.46$

\item[Fig.15] The relaxation times $\tau$ 
versus antiferromagnetic interactions concentration $\kappa$ at
$T=T_c^0(\kappa)$ for the Ising Model with variable antiferromagnetic
interactions of size $L=32$. $T=T_c^0(\kappa)$ is defined as the temperature at
which the susceptibility of a system of size $L=32$ gets its maximum.
Two MC dynamics are compared: unmodified FK-CK cluster dynamics
(squares) and the cluster dynamics introduced in section VIII.C
(triangles).

\item[Fig.16] The relaxation function, $q$ (see eq.79), for Ising SG model
versus time (MC Steps per spin). The temperature reported are, from the bottom
to the top, $T=1.4,1.3,1.2,1.1,1.0,0.9,0.8$ and 
the system sizes are: $L=80,90,100$.
The cluster dynamics used is described in sections VIII.C.

\end{enumerate}


\begin{thebibliography}{9999}
\bibitem{nota_a} An interaction of 
strength $J'$ which eventually will be put equal to $\infty$ is also called 
bond.
\bibitem{Notazza} In the case of a deterministic interaction configuration 
$T_c$ is obviously defined without the average over the 
interaction configuration.
\bibitem{nota} The restriction to a square lattice does not imply any lost of 
generality. The arguments given are valid for any lattice.
\bibitem{note} Some results contained in this section have been reported in 
Ref.[11].
\bibitem{nota_equiv} If we used eqs.(\ref{55}) instead of
(\ref{condConiglioblockfond}) the solutions $w_7=w_8=w_9=0$
($w_5=w_7=w_9=0$) for $u>u^*$ ($u<u^*$) would be found immediately. Then
the not zero weights could be calculated by using eqs.(\ref{set_1}) only.
\bibitem{mistake} Our preliminary numerical analysis appeared in 
Ref.\cite{letter}
indicated $\gamma_p(1)\simeq 7/4$ and $\nu_p(1)\simeq 1$.
\end{thebibliography}
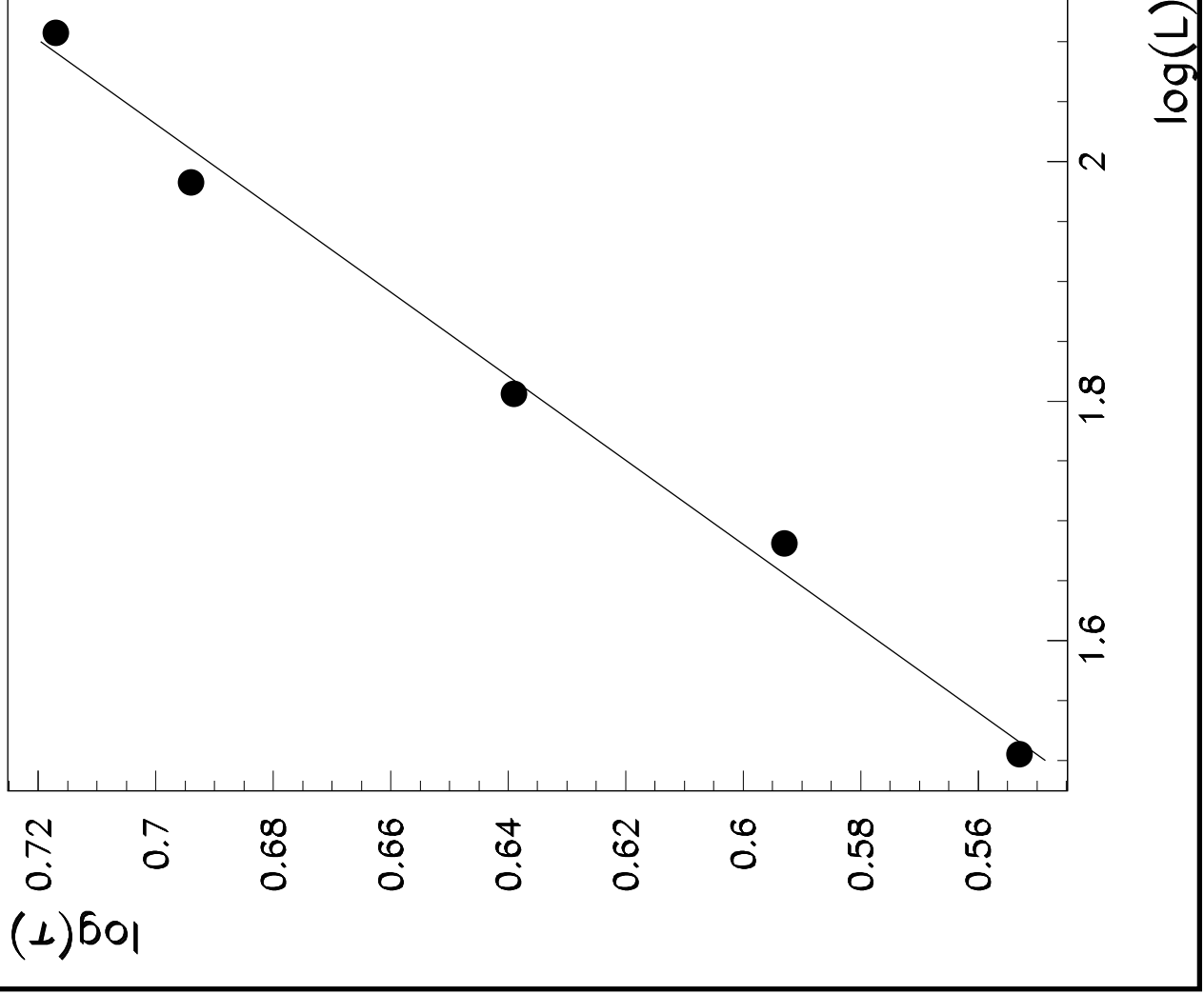
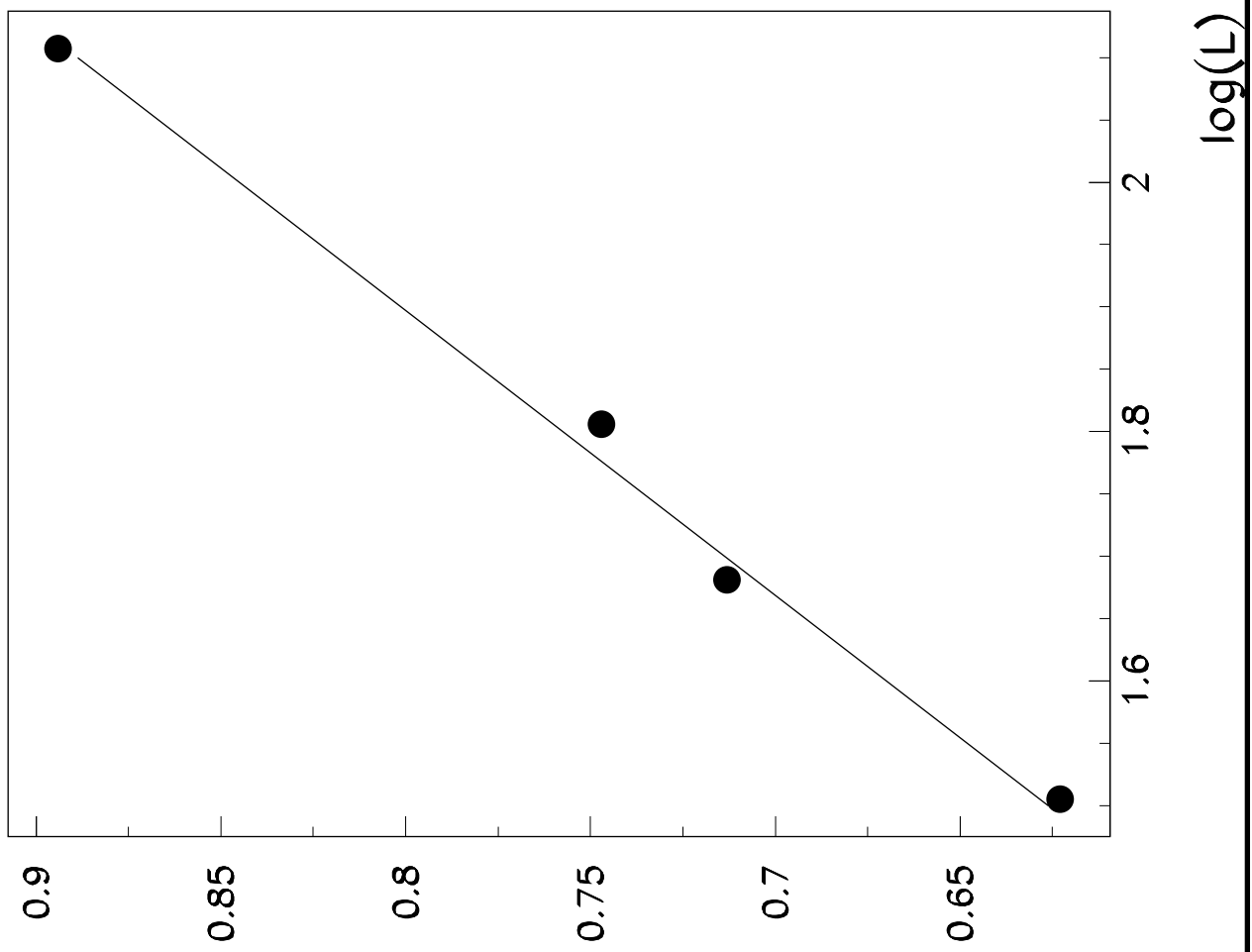
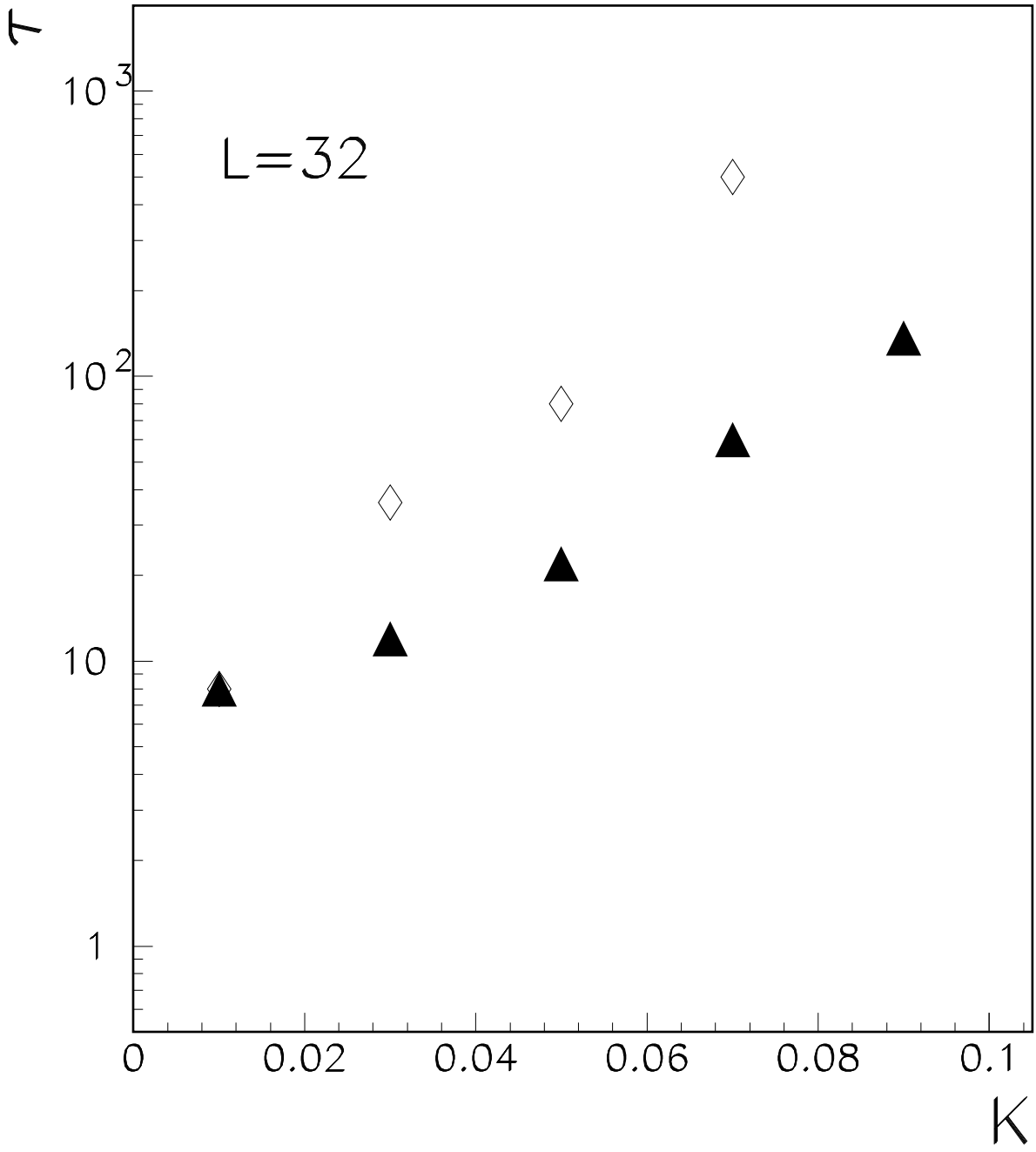
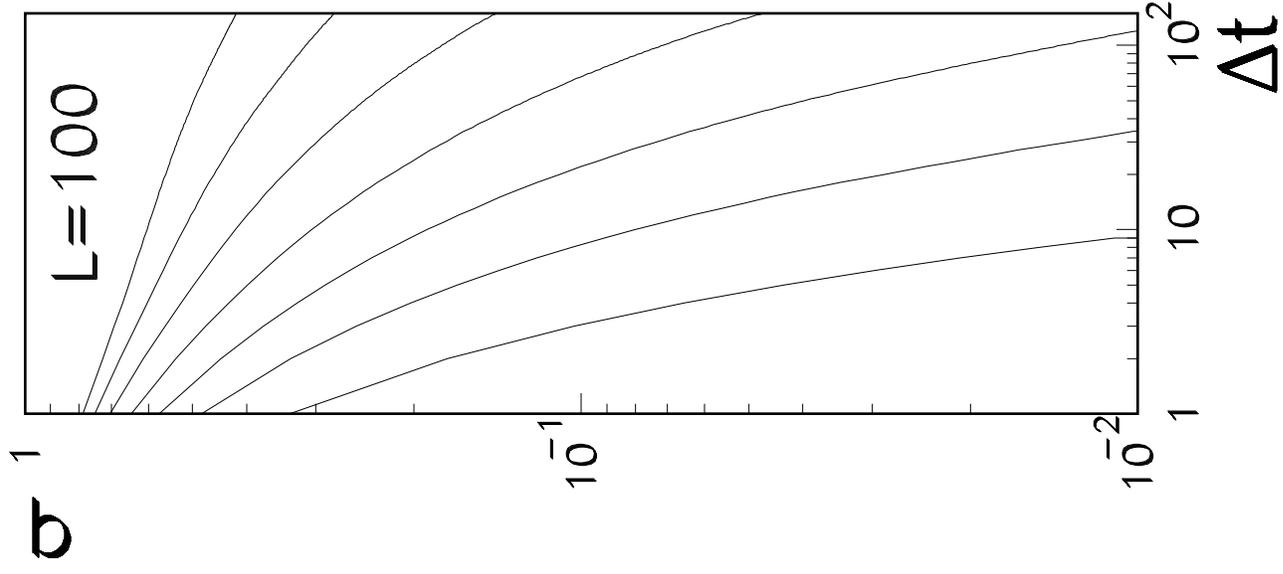
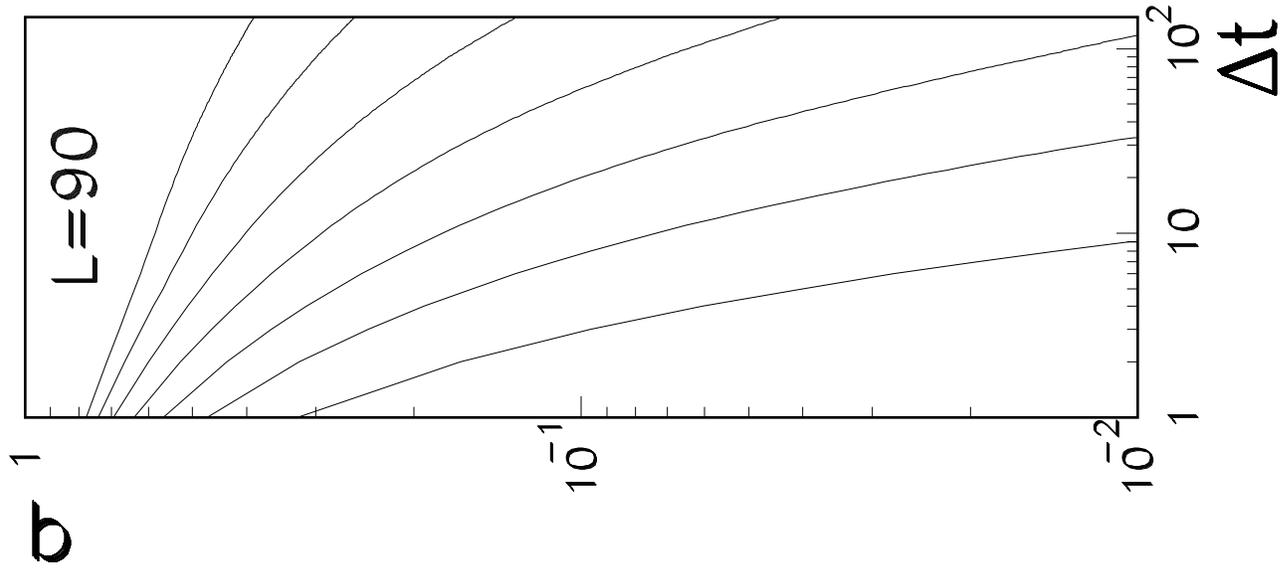
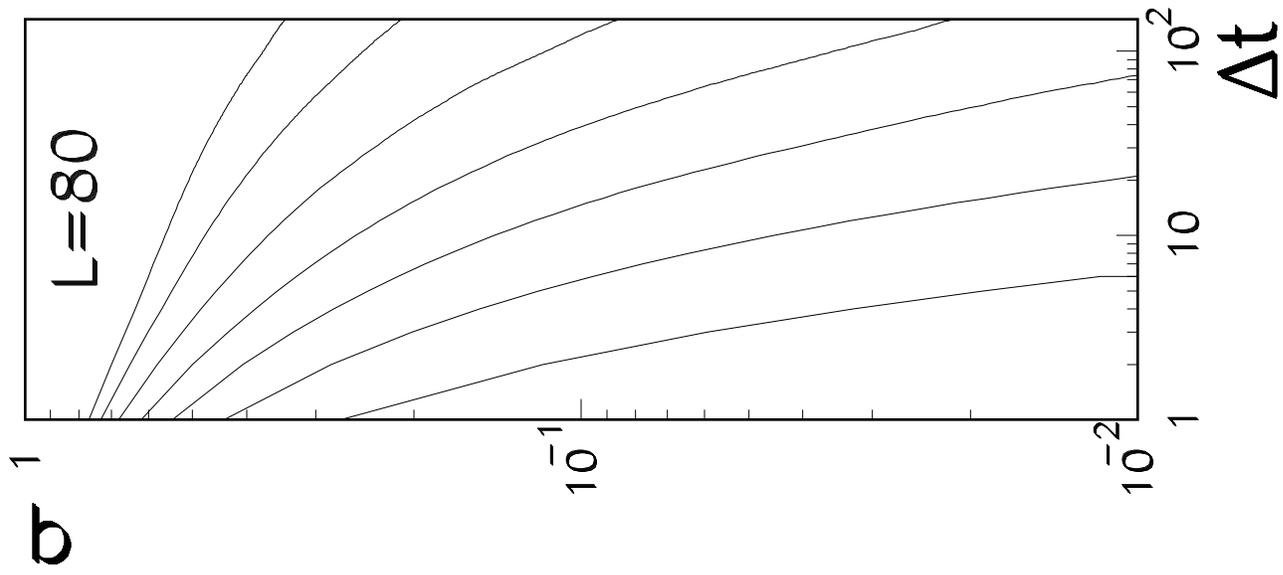
\end{document}